# Compressive single-pixel imaging via a wavelength-multiplexed spatially incoherent diffractive optical processor

Xiao Wang[1,2,3], Yiyang Wu[1,2,3], Yuntian Wang[1,2,3], Md Sadman Sakib Rahman[1,2,3], Paloma Casteleiro Costa[1,2,3], Guangdong Ma[1,2,3], Shiqi Chen[1,2,3], Yuzhu Li[1,2,3], Jingxi Li[1,2,3], Çağatay Işıl[1,2,3] and Aydogan Ozcan[1,2,3,*]

[1]Electrical and Computer Engineering Department, University of California, Los Angeles, CA, 90095, USA
[2]Bioengineering Department, University of California, Los Angeles, CA, 90095, USA
[3]California NanoSystems Institute (CNSI), University of California, Los Angeles, CA, 90095, USA
[*]ozcan@ucla.edu

## Abstract

Despite offering high sensitivity, a high signal-to-noise ratio, and a broad spectral range, single-pixel imaging (SPI) is limited by low measurement efficiency and long data-acquisition times. To address this, we propose a wavelength-multiplexed, spatially incoherent diffractive optical processor combined with a compact/shallow digital artificial neural network (ANN) to implement compressive SPI. Specifically, we model the bucket detection process in conventional SPI as a linear intensity transformation with spatially and spectrally varying point-spread functions. This transformation matrix is treated as a learnable parameter and jointly optimized with a shallow digital ANN composed of 2 hidden nonlinear layers. The wavelength-multiplexed diffractive processor is then configured via data-free optimization to approximate this pre-trained transformation matrix; after this optimization, the diffractive processor remains static/fixed. Upon multi-wavelength illumination and diffractive modulation, the target spatial information of the input object is spectrally encoded. A single-pixel detector captures the output spectral power at each illumination band, which is then rapidly decoded by the jointly trained digital ANN to reconstruct the input image. In addition to our numerical analyses demonstrating the feasibility of this approach, we experimentally validated its proof-of-concept using an array of light-emitting diodes (LEDs). Overall, this work demonstrates a computational imaging framework for compressive SPI that can be useful in applications such as biomedical imaging, autonomous devices, and remote sensing.

# Introduction

Unlike conventional sensor arrays featuring millions of pixels, single-pixel imaging (SPI) employs a single-pixel detector to form spatially resolved images[1–4]. Owing to the advantages of single-pixel detectors, including high sensitivity, low noise, broad spectral applicability, short response time, and lower costs, SPI has been widely applied in spectral imaging[5–7], terahertz imaging[1,8–11], X-ray imaging[12–14], remote sensing[15,16], target tracking[17–19], three-dimensional imaging[6,20,21], photoacoustic imaging[22–24], complex-amplitude light-field reconstruction[25–28], and imaging through scattering media[29–31]. In traditional implementations, SPI encodes a target with a set of modulation patterns applied actively or passively, and the resulting integrated bucket signals are subsequently decoded via specialized algorithms to reconstruct the image[4]. Although the benefits of single-pixel detection have been successfully demonstrated in many fields, this approach inevitably entails a time-consuming measurement process, which partially limits its broader applicability. To alleviate this bottleneck, researchers have sought to replace random speckles with more effective modulation patterns—either optimized[32], orthogonal[33–36] or learned[37–39]—in order to reduce the number of required measurements. At the same time, more advanced reconstruction algorithms such as differential ghost imaging[40], alternating projections[41], compressive sensing[42,43], and deep learning-based algorithms[44–46] have been introduced to achieve higher-quality reconstructions from fewer measurements. Nevertheless, an important challenge remains: these efficient modulation patterns often rely on frame-by-frame display via, e.g., spatial light modulators (SLMs), which inherently dictate and limit the maximum achievable acquisition speed.

Diffractive deep neural networks (D²NNs) have recently emerged as a powerful optical computing framework[47], opening new avenues for optical imaging, sensing and information processing. The central concept lies in the analog processing of optical fields through cascaded diffractive layers, each comprising tens to hundreds of thousands of trainable diffractive features. Through a deep learning-based training phase, the transmission coefficients of these diffractive elements are optimized to spatially modulate the incident light, accurately performing a desired function between the input and output apertures. The D²NN architecture performs computational tasks through passive optical propagation, while offering benefits in parallelism, processing speed, and energy efficiency compared with conventional electronic implementations. D²NNs have been successfully applied to object classification[47–54], nonlinear function approximation[55,56], quantitative phase imaging[57–59], holographic reconstruction[60], logic operations[61–63], and universal linear transformations[64–67]. Further research has extended D²NNs to broadband optical field modulation, achieving applications such as pulse shaping[68], wavelength-division multiplexing[69], and single-pixel machine vision and sensing[70]. At the intersection of deep learning and optical design, D²NNs have evolved into a highly scalable, all-optical architecture for high-throughput, low-power computational imaging and sensing systems.

Here, we demonstrate compressive SPI utilizing a wavelength-multiplexed spatially incoherent diffractive optical processor. Specifically, the sequential measurement process in conventional SPI can be modeled as a linear transformation linking a target object to its detected single-pixel intensities. We use a diffractive neural network equipped with a sufficient number of trainable diffractive features to approximate this desired linear transformation using spatially- and spectrally-varying point-spread function engineering. To implement this, we first embed a randomly initialized linear transformation matrix and feed it into a shallow digital artificial neural network (ANN) with 2 hidden layers. By jointly training this SPI architecture, we obtain a target linear transformation matrix alongside an optimized decoder back-end ANN. Next, using a data-free optimization approach combined with wavelength multiplexing, we design an incoherent diffractive optical processor that approximates the learned target linear intensity transformation under spatially incoherent illumination composed of a predetermined set of illumination wavelengths. The shallow decoder ANN is then fine-tuned using the D²NN-estimated linear transformation matrix to address the performance gap caused by the diffractive network approximation; this transfer learning step effectively bridges the numerical discrepancies between the digital twin and the multi-wavelength, spatially incoherent D²NN implementation. With this learning process, the need for sequentially loading modulation patterns is replaced by the propagation and diffraction of the illumination light at a set of pre-determined wavelengths through a static/fixed diffractive optical processor. By removing the need for SLM-based modulation, this diffractive optical architecture enables measurement speeds primarily determined by the single-pixel detector readout/refresh rate. Notably, this wavelength-multiplexed framework can be extended to various parts of the electromagnetic spectrum by appropriately scaling the physical dimensions of the diffractive features in proportion to the operating wavelength. We believe this resource-efficient, low-latency, and low-power approach for spectral target encoding may provide a useful framework for future machine vision applications.

## Results

Throughout this work, the terms "diffractive deep neural network", "diffractive neural network", and "diffractive processor/network" are used interchangeably to refer to the same class of free-space optical computing architectures. Conventional SPI is a computational imaging method that reconstructs the spatial distribution of a target using a single-pixel detector. As shown in **Fig. 1a**, its fundamental principle relies on employing a sequence of known modulation patterns (using, e.g., an SLM) to encode the target object. In each measurement, only a single intensity value is recorded. By performing multiple measurements and applying numerical reconstruction algorithms, the target image can be recovered. Mathematically, the SPI process can be abstracted as a linear transformation. Assuming the target image is represented by a vector $\boldsymbol{i} \in \mathbb{R}_+^{N_i}$, the entire set of modulation patterns can be formulated as a sensing matrix $\mathbf{A} \in \mathbb{R}_+^{N_o \times N_i}$. In this matrix, each row dictates a specific spatial modulation pattern, with $N_i$ representing the total number of pixels in the input image and $N_o$ denoting the total number of sequential measurements. The resulting bucket measurement vector $\boldsymbol{o} \in \mathbb{R}_+^{N_o}$ is thus given by:

$$\boldsymbol{o} = \mathbf{A}\boldsymbol{i} + \boldsymbol{n}, \tag{1}$$

where $\boldsymbol{n}$ denotes the noise term. This formulation reveals that the sequential modulation process in SPI essentially projects the high-dimensional image signal $\boldsymbol{i}$ onto a lower-dimensional measurement space. The design of the sensing matrix $\mathbf{A}$, the compression ratio $\gamma$ ($\gamma = \frac{N_o}{N_i}$), and the reconstruction algorithm jointly dictate the ultimate image reconstruction quality.

In this work, we employ a static/fixed wavelength-multiplexed incoherent diffractive processor with spatially and spectrally varying point-spread functions to implement the linear transformation corresponding to the matrix $\mathbf{A}$ in **Eq. (1)** without the need for sequential updates on an SLM. The basic architecture of this multi-wavelength diffractive processor is illustrated in **Fig. 1b**. Specifically, a static, broadband diffractive neural network with a single diffractive layer is structurally optimized to approximate the target, multi-wavelength linear transformation obtained in a pre-training stage. In the example shown in **Fig. 1b**, the transformation matrix $\mathbf{A}$ has dimensions of $32 \times 64$, corresponding to a compression ratio of $\gamma = 32/64 = 0.5$. A target object with a size of $8 \times 8$ pixels is illuminated by spatially incoherent light, and encoded by the diffractive processor into $N_w = 32$ wavelength-dependent intensity channels with a pre-determined set of illumination wavelengths. These wavelength-encoded signals are captured by a single-pixel detector and subsequently decoded by a shallow digital ANN with 2 hidden layers for image reconstruction. Using a spectrum analyzer, the output intensity values at all 32 wavelength channels can be measured simultaneously; alternatively, the illumination wavelengths can be switched sequentially. In our numerical analysis, without loss of generality, the 32 selected wavelengths are uniformly distributed within the range of 2.7 mm to 4.0 mm, yielding a mean wavelength of $\lambda_m = 3.35$ mm. The lateral size of the input target is ~16 $\lambda_m$, while the lateral size of the single pixel at the output plane is ~5 $\lambda_m$. Additionally, the distance from the target to the diffractive layer is ~36 $\lambda_m$, and ~96 $\lambda_m$ from the diffractive layer to the detection plane. The diffractive layer consists of $64 \times 64$ trainable diffractive features, each laterally sized at ~$\lambda_m/2$. Without the need for redesigning its features, the same architecture (after its optimization) can operate at other parts of the electromagnetic spectrum by scaling the diffractive features proportional to $\lambda_m$.

As illustrated in **Fig. 1c**, the training of the multi-wavelength spatially incoherent diffractive processor consists of the following 3 steps. First, the target linear transformation **A** (which in general corresponds to a fully connected linear intensity transformation) is fed into a shallow ANN with 2 hidden layers. In this digital twin, both the linear transformation matrix **A** and the ANN are randomly initialized. We then jointly optimize the linear transformation matrix **A** and the ANN ($\mathcal{R}_\theta$) using the following objective function (loss function):

$$\{\mathcal{R}_{\theta^*}, \mathbf{A}^*\} = \mathrm{argmin}_{\theta,\mathbf{A}} \|\mathcal{R}_\theta(\mathbf{A}i) - i\|^2, \tag{2}$$

where $\theta$ denotes the learnable parameters of the digital ANN. The optimized linear transformation matrix $\mathbf{A}^*$ is regarded as the target linear transformation matrix, i.e., $\mathbf{A}_\mathrm{target} = \mathbf{A}^*$. In the second step, the multi-wavelength spatially incoherent diffractive processor is structurally optimized to approximate the target transformation, $\mathbf{A}_\mathrm{target}$. For this task, we adopt a *data-free* D²NN optimization method (see **Methods** for more details). This method optimizes the trainable phase profile of the diffractive layer by propagating each input pixel separately, and minimizes the loss function, which is given by:

$$\mathcal{D}_{h^*} = \mathrm{argmin}_h \|\mathcal{D}_h(\mathbf{U}) - \mathbf{A}_\mathrm{target}\|^2 + \alpha \mathcal{L}_\mathrm{eff}, \tag{3}$$

where $\mathcal{D}_h$ represents the forward operator of the spatially incoherent multi-wavelength diffractive processor, $h$ denotes the trainable parameters of the diffractive layer, **U** is the unit-intensity matrix, $\mathcal{L}_\mathrm{eff}$ is the output diffraction efficiency-related penalty term, and the hyperparameter $\alpha$ is introduced to balance the relative contributions of these loss terms during training. The diffraction efficiency penalty term in **Eq. (3)** is defined as:

$$\mathcal{L}_\mathrm{eff} = \max\left(0, -\log\left(\frac{\eta}{\eta_\mathrm{th}}\right)\right), \tag{4}$$

where $\eta$ denotes output diffraction efficiency and $\eta_\mathrm{th}$ represents the diffraction efficiency threshold (which is a design hyperparameter), and $\eta$ is defined as:

$$\eta = \frac{\langle P_o \rangle_\lambda}{P_i}, \tag{5}$$

where $P_o$ denotes the single-pixel output power of the diffractive processor, $P_i$ denotes the input power, and $\langle \cdot \rangle_\lambda$ denotes the ensemble average over the wavelength-multiplexed channels. After optimization, the approximated transformation matrix of the diffractive processor is obtained as $\mathbf{A}_\mathrm{estimated} = D_{h^*}(\mathbf{U})$. Compared to data-driven learning approaches, the data-free D²NN optimization neither requires large training datasets nor extensive simulations of incoherent light propagation, thereby greatly improving the efficiency of optimizing the spatially incoherent diffractive processor. Finally, the approximated matrix $\mathbf{A}_\mathrm{estimated}$ is used to replace $\mathbf{A}_\mathrm{target}$ obtained in the first step, and its parameters are *frozen*. The shallow digital ANN $\mathcal{R}_{\theta^*}$ is then further fine-tuned as follows:

$$\mathcal{R}_{\theta^{**}} = \mathrm{argmin}_{\theta^*} \|\mathcal{R}_{\theta^*}(\mathbf{A}_\mathrm{estimated} i) - i\|^2. \tag{6}$$

Through this fine-tuning stage, the shallow digital ANN becomes better adapted to the optical diffractive processor, resulting in higher-quality decoding and output image reconstruction. This step essentially corrects for numerical discrepancies that are encountered between the digital twin and the diffractive optical network learning processes

- mostly driven by the additional diffraction efficiency-related loss term, $\mathcal{L}_{\text{eff}}$, in **Eq. (3)**, that does not appear in our digital twin.

Next, we present numerical simulation results to evaluate the performance of the proposed wavelength-multiplexed incoherent diffractive processor. In all numerical simulations, we used the STL10 dataset[71] for training. This dataset contains 100,000 natural images spanning 10 categories, each with an original resolution of 96 × 96 pixels, which were downsampled to 8 × 8 for our implementation. The data was split into 90,000 training images, 9,000 validation images, and 4 test images, randomly selected from the remaining 1,000 images. To further evaluate the external generalization capabilities of our model, in addition to the 4 STL10 test images, we constructed an additional test set of 46 images, including handcrafted digits, letters, resolution patterns, human faces[72], and other images, for a total of 50 test samples never seen during our training (see **Supplementary Fig. S1** for details). It is important to emphasize that the training data were used only for training (step 1) and fine-tuning (step 3) of the digital ANN, while the diffractive neural network was optimized using a data-free approach. Furthermore, to maintain acceptable diffraction efficiency, we set the output diffraction efficiency threshold $\eta_{\text{th}}$ to 0.5% and $\alpha =$ 0.1 during the optimization.

As shown in **Fig. 2a1**, after optimization, the diffractive network encodes the test images into 32-wavelength power spectra, which are captured by a single-pixel detector and reconstructed by the jointly trained shallow digital ANN. **Fig. 2a2** illustrates the phase distribution of the diffractive layer obtained through the data-free D²NN optimization. **Fig. 2a3** compares the target linear transformation with the estimated one, achieving a cosine similarity (CosSim) of 0.93, a mean squared error (MSE) of 0.04, and a peak signal-to-noise ratio (PSNR) of 13.50 dB (see **Methods** for details on the quantitative metrics). We note that the learning constraint on target diffraction efficiency (i.e., $\mathcal{L}_{\text{eff}}$) is a primary source of the small deviations observed between the estimated and target transformations. Nevertheless, by fine-tuning the back-end digital ANN with the estimated transformation in step 3, the model maintains stable image reconstruction performance.

**Fig. 2b** presents representative imaging results, structured in four columns corresponding to the ground truth images, single-pixel spectral measurements, reconstructed images, and the error maps between the ground truth images and the reconstructions. While these tests were conducted on 50 test images never seen during training, 12 representative examples are displayed here, with sample indices denoted below the ground truth images. The selected targets include digits, resolution test patterns, letters "U," "C," "L," and "A" against both black and white backgrounds, and other images, which were never used during training. Visually, the reconstructed images maintain high consistency with their respective ground truths across varying data distributions, indicating the reliable generalization capability of our model. Quantitatively, evaluating all 50 test samples yielded an average MSE of 0.04 ± 0.03, PSNR of 15.82 ± 4.51 dB, and structural similarity index (SSIM) of 0.83 ± 0.11. **Fig. 2c** further shows bar plots of the quantitative metrics for the 12 representative samples. Notably, because the model was trained on natural images from the STL10 dataset[71], better reconstruction quality is observed for samples 9–12, which exhibit continuous and uniform grayscale distributions. Although a slight degradation in reconstruction quality is noted for the other out-of-distribution image data categories, the model consistently maintains satisfactory reconstruction performance. Overall, these findings demonstrate the diffractive SPI architecture's broad generalization capacity across diverse imaging targets.

Next, we investigate the impact of various hyperparameters on model performance. Specifically, we evaluate how the number of diffractive layers, single-pixel width ($w$), output diffraction efficiency ($\eta$), diffractive layer phase bit depth, and the number of multiplexed wavelengths influence both the network's ability to approximate the target linear transformation and the final image reconstruction quality. To quantify the fidelity of the linear transformation, we compute the MSE, PSNR, and CosSim metrics between the target linear transformation $\mathbf{A}_{\text{target}}$ and the estimated transformation $\mathbf{A}_{\text{estimated}}$. Correspondingly, the final image reconstruction performance is assessed using MSE, PSNR, and SSIM metrics. Notably, we evaluate three distinct architectural configurations comprising K = 1, 2, and 3 diffractive layers—each maintaining a spatial resolution of 64 × 64 pixels. These configurations are analyzed in conjunction with the other four hyperparameters by isolating the effect of a single variable while holding all others constant. In this analysis, we first report the impact of the single-pixel/detector width ($w$) on model performance. In our simulations, we considered four different detector widths ($w$): 13.6 mm ($\sim 4\lambda_m$), 17.0 mm ($\sim 5\lambda_m$), 20.4 mm ($\sim 6\lambda_m$), and 23.8 mm ($\sim 7\lambda_m$). The other hyperparameters were fixed as follows: output diffraction efficiency threshold, $\eta_{th} = 0.5\%$, diffractive layer phase bit depth of 8 bits, and $N_w = 32$. As depicted in **Fig. 3a**, the top row illustrates the diffractive network's approximation of the target linear transformation, while the bottom row shows the final image reconstruction performance. As $w$ increases, the similarity between the estimated and target linear transformations improves, and the error decreases, although the improvement trend gradually saturates. This suggests that, under the optimization constraint of a target diffraction efficiency (see **Eq. (3)**), the diffractive neural network sacrifices some performance to maintain a sufficient output diffraction efficiency, and larger detector sizes help alleviate this trade-off. However, the improvement is limited, and further increasing the detector size does not yield additional performance gains. This behavior remains consistent across varying numbers of diffractive layers (K = 1, 2, 3). As expected, the deeper, three-layer diffractive network configuration achieves the highest linear transformation accuracy.

Intuitively, a closer optical approximation of the target transformation by an optimized diffractive network is expected to improve the image decoding process of the shallow digital ANN. Indeed, the imaging performance curves in **Fig. 3a** exhibit a trend similar to that of the linear transformation performance. However, a comparison of the ANN decoding performance between the K = 3 and K = 2 configurations reveals a counterintuitive behavior: although the three-layer diffractive processor—equipped with more trainable diffractive elements—achieves a more accurate approximation of the target linear transformation, the resulting image quality decoded by the shallow ANN shows no significant advantage over the two-layer configuration. Two factors may help explain this behavior. First, the target transformation matrix $\mathbf{A}_{\text{target}}$, obtained through joint training, does not necessarily represent the globally optimal sensing matrix. Thus, the error observed in the optically implemented $\mathbf{A}_{\text{estimated}}$ compared to $\mathbf{A}_{\text{target}}$ does not strictly equate to a proportional loss in the image reconstruction quality. Second, the digital ANN is explicitly fine-tuned against errors introduced in $\mathbf{A}_{\text{estimated}}$ to optimize the final image reconstruction step. Therefore, the robust computational decoding ability of the shallow digital back-end effectively compensates for the suboptimal encoding at the optical stage. As a result, the final imaging performance depends on the combined effects of optical encoding fidelity and digital error compensation.

Diffraction efficiency ($\eta$) is another critical metric for diffractive optical neural networks, as it is closely related to their practical applicability. Two main factors affect the output diffraction efficiency: (i) the intrinsic properties of the diffractive material, whereby part of the optical energy is absorbed during light–matter interactions; and (ii) propagation losses, where part of the optical energy leaks out of the diffractive processor's computing volume. The latter can be mitigated by incorporating a diffraction-efficiency penalty term into the training loss function; see **Eq. (3)**. To shed more light on this, while keeping the single-pixel width ($w$) fixed at $\sim 7\lambda_m$, the diffractive layer phase bit depth at 8 bits, and $N_w$ at 32, we investigated the impact of four diffraction efficiency thresholds, $\eta_{th}$: 0.3%, 0.5%, 0.7%, and 1%. The results, shown in **Fig. 3b**, indicate that as the diffraction efficiency threshold increases, demanding an increased $\eta$, the associated penalty term becomes stronger, leading to higher errors between the estimated and target linear transformations, with imaging quality exhibiting a similar trend. These findings suggest that the diffractive model tends to sacrifice some performance to maintain diffraction efficiency. Therefore, in practical applications, a careful trade-off between the model performance and diffraction efficiency should be considered to achieve an optimal configuration tailored to the intended use cases.

Next, we assessed the impact of phase quantization on model performance by evaluating linear transformation accuracy and image reconstruction quality across five phase bit depths: 1, 2, 4, 8, and 12 bits. For this analysis, we held the single-pixel width ($w$) at $\sim 7\lambda_m$, the output diffraction efficiency threshold ($\eta_{th}$) at 0.5%, and the number of multiplexed wavelengths ($N_w$) at 32. As shown in **Fig. 4a**, the 8-bit and 12-bit diffractive processors exhibited comparable accuracy in optical linear transformation and image reconstruction performance, while the 4-bit configuration showed only a slight degradation. This indicates that in scenarios where high bit-depth diffractive surface fabrication is difficult to achieve, it is feasible to adopt 4-bit manufacturing standards, with limited performance degradation.

The number of multiplexed wavelengths ($N_w$) dictates the spectral channels used to encode the target object's spatial information. For a fixed target size of $8 \times 8 = 64$ pixels, introducing more illumination wavelengths creates a higher-dimensional sensing matrix and increases the sampling rate. We tested 16, 32, and 48 multiplexed wavelengths, keeping the single-pixel width ($w$) at $\sim 7\lambda_m$, output diffraction efficiency threshold ($\eta_{th}$) at 0.5%, and the diffractive layer phase bit depth at 8 bits. **Fig. 4b** presents the model performance in terms of both linear transformation accuracy and image reconstruction quality under different numbers of multiplexed wavelengths. The accuracy of the optical linear transformation notably declines with more wavelengths, as a fixed number of diffractive elements lacks the degrees of freedom required to accurately map a substantially larger, more complex transformation matrix represented by spectral diversity (a larger $N_w$). Conversely, the final image reconstruction quality steadily improves with increasing $N_w$. These results suggest a benefit of the hybrid framework: the increased measurement information provided by a higher sampling rate (larger $N_w$), together with the digital ANN's decoding capability, can effectively offset the performance loss caused by the degraded diffractive encoding step.

Next, we perform a proof-of-concept experiment in the visible spectrum to validate the proposed wavelength-multiplexed spatially incoherent diffractive processor. Using the experimental setup illustrated in **Fig. 5a, b** (see **Methods** for details), we demonstrate compressive SPI on target images sized at 3 × 3 pixels, using 5 illumination wavelengths. This corresponds to a compression ratio of $\gamma = 5/9 \approx 0.56$. To generate the multi-wavelength incoherent illumination, five LEDs with central wavelengths of 505 nm, 523 nm, 605 nm, 623 nm, and 660 nm were arranged in a ring configuration. During the measurement, the LEDs were sequentially turned on one at a time. The emitted light was collimated to illuminate a digital micromirror device (DMD) displaying the target objects. The DMD surface, i.e., the object plane, was then conjugated to an intermediate image plane via a 4f system. From there, the light propagated a distance of d1 = 5 mm, underwent modulation by the single-layer diffractive processor, and propagated d2 = 12.4 mm to a complementary metal-oxide-semiconductor (CMOS) imager. A 7 × 7-pixel region was cropped and integrated from the CMOS imager to act as the wavelength-dependent single-pixel intensity signal.

The diffractive processor comprising a single diffractive layer was optimized using the same methodology as in our numerical simulations. The optimized thickness profile of the diffractive layer is depicted in **Fig. 5c**, while the physical layer, fabricated using a two-photon polymerization-based 3D printer, is shown in **Fig. 5d-f** (see **Methods** for fabrication details). To ensure alignment during the experiment, 3D translation stages were utilized to finely position both the diffractive layer and the detector relative to the incident beam. For the fine-tuning of the back-end digital ANN in step 3, we utilized an experimentally collected dataset comprising 512 pairs of 5-wavelength multiplexed intensity measurements and their corresponding ground-truth target images. This dataset consists of all possible configurations of 3×3 binary images ($2^9 = 512$ image patterns), and it is partitioned into 500 training samples, 6 validation samples, and 6 test samples, with the test set completely excluded during both the training and validation stages. The imaging results for the test set are presented in **Fig. 5g**. The first column displays the ground-truth targets, while the second column compares the normalized simulated and experimental wavelength-multiplexed single-pixel intensity signals. The third and fourth columns show the numerically simulated and experimentally measured image reconstructions achieved by the diffractive processor, respectively. For comparison, we also conducted two additional control experiments while keeping all other system components unchanged: removing the diffractive layer entirely (i.e., free-space propagation) and replacing it with a random diffuser; in each control experiment, the back-end ANN was separately trained using experimental data to provide a fair comparison. The experimental reconstructions for these two baselines are shown in the fifth and sixth columns of **Fig. 5g**. The quantitative image quality metrics for these four scenarios are also summarized in **Table 1**. Compared to both free-space propagation and the random diffuser, the diffractive processor achieves better imaging quality in both simulations and experiments, confirming the efficacy of the optical compressive encoding realized by the diffractive layer.

Notably, an observable mismatch exists between the simulated and experimental wavelength-encoded signals (**Fig. 5g**, column 2). This discrepancy may arise from two main factors. First, practical imperfections during the lithographic 3D fabrication process of the physical diffractive layer introduce structural deviations from the ideal simulated profile. Second, given the sub-micron wavelengths in the visible spectrum, even with the aid of 3D translation stages, any minor lateral or axial misalignments—along with imperfections in the 4f system—can introduce deviations between the experimental and simulated results. However, it must be emphasized that our proposed paradigm—combining optical diffractive encoding with digital decoding—helps reduce the impact of these physical errors. Specifically, the shallow digital ANN, initially trained *in silico*, undergoes transfer learning on the experimentally acquired dataset to account for and adapt to these systemic errors, thereby enabling the decoder ANN to generalize robustly to the actual experimental environment. As shown in **Fig. 5g** and **Table 1**, despite the distorted raw optical signals, the final experimental imaging quality shows limited degradation relative to the corresponding simulation results.

## Discussion

In this work, we utilize a wavelength-multiplexed spatially incoherent diffractive processor to execute an optical linear transformation, conceptually mirroring the sequential modulation of conventional compressive SPI but encoding spatial information in the spectral domain. Through joint pre-training with a shallow digital decoder ANN, this target linear transformation is optimized as learnable parameters—a process analogous to learning highly efficient modulation patterns in SPI[37,39]. Consequently, this learned transformation defines a spatial-to-spectral encoding strategy. Following this, we configure the multi-wavelength diffractive processor to optically approximate the target transformation using a *data-free* D²NN optimization method. Notably, simulating spatially incoherent illumination within a standard data-driven training pipeline for a diffractive processor requires combining each input sample with large ensembles of random illumination phases, resulting in a high computational cost. By employing a data-free D²NN optimization approach, we effectively bypass this limitation and achieve efficient optimization. The resulting optically estimated transformation deviates slightly from the target transformation, mostly due to the diffraction efficiency-related loss term used during optimization; however, we address this by fine-tuning the digital ANN using the estimated optical transformation. In this way, the optical-digital co-design improves the ability of the digital back-end to decode the spectral signals measured from the diffractive processor.

By optically realizing the linear transformation of sequential SPI and encoding the data into the spectral domain, the imaging speed is no longer limited by the SLM refresh rate. Instead, it is governed by the modulation rate of the light source and the sampling rate of the single-pixel detector. Single-pixel detectors based on photodiodes or photomultiplier tubes can readily achieve sampling rates in the MHz–GHz range, while the wavelength modulation of the illumination can be implemented using LED arrays (or laser diodes in the coherent case). Moreover, because the static diffractive processor operates through optical diffraction, it can be adapted to various regions of the electromagnetic spectrum. Extension to other wavelengths may be achieved by appropriately scaling the diffractive feature sizes, without the need for retraining.

In our numerical analysis, the baseline model employed 32 multiplexed wavelengths, and we further evaluated its performance with 16 and 48 wavelengths. We note that this setting should not be interpreted as the maximum number of spectral channels supported by the model. Previous studies have demonstrated the ability of diffractive networks to process optical waves across large frequency ranges[69]. More recently, Li *et al*. reported on the robust information processing capabilities of broadband diffractive networks, revealing their ability to handle ~2000 spectral channels while maintaining high computational fidelity[65].

Despite its high efficiency, speed, and low power consumption, the proposed incoherent diffractive processor also faces some scaling challenges. Imaging large field-of-view scenes at high resolution generates a large-scale target linear transformation matrix, even under high compression conditions. Consequently, this necessitates (i) more diffractive elements or layers to optically realize the target transformation, which substantially increases optimization complexity and computational cost, and (ii) denser wavelength multiplexing. Within a limited spectral range, smaller wavelength spacing can induce performance crosstalk, reducing the diffractive processor's accuracy. Nevertheless, future improvements in computational resources and better 3D fabrication methods, together with the multi-wavelength adaptability of the proposed incoherent diffractive processor, may help mitigate some of these challenges.

In conclusion, we presented a wavelength-multiplexed spatially incoherent diffractive processor designed to replace the sequential modulation of conventional SPI with an optical, spectrally encoded linear transformation. By coupling this optical front-end with a shallow digital ANN for image reconstruction, our hybrid framework realizes resource-efficient spatial-to-spectral encoding with low latency and power consumption. This method aims to circumvent the acquisition bottlenecks inherent to traditional SPI, paving the way for faster compressive imaging. Moreover, its diffraction-based physical scaling allows it to be adapted across diverse spectral bands. We believe that the proposed framework may hold promise for applications such as biomedical imaging, autonomous driving, and remote sensing.

## Methods

### Data-free D²NN optimization

As formulated in **Eq. (3)**, we employ a data-free D²NN optimization scheme similar to the approach detailed in our previous implementation[66]. Specifically, this optimization is achieved by minimizing the MSE between the estimated all-optical intensity transformation $\mathbf{A}_{\text{estimated}}$ executed by the spatially incoherent diffractive network and the target transformation $\mathbf{A}_{\text{target}}$. To computationally evaluate the $\mathbf{A}_{\text{estimated}}$, we utilize a set of $N_i$ input intensity vectors, denoted as $\{\boldsymbol{i}_t\}_{t=1}^{N_i}$. For any given vector $\boldsymbol{i}_t$, its $l$-th element is defined such that $\boldsymbol{i}_t[l] = 1$ when $l = t$, and 0 otherwise. In other words, the set $\{\boldsymbol{i}_t\}_{t=1}^{N_i}$ represents a series of unit impulse functions spatially shifted across the various input pixels, which can be represented in matrix form as $\mathbf{U}$. We simulate the corresponding all-optical output intensity vectors $\{\boldsymbol{o}_t\}_{t=1}^{N_i}$, which consist of the optical responses distributed across the different multiplexed wavelength channels. These output vectors are subsequently concatenated column-wise to construct the estimated transformation matrix:

$$\mathbf{A}_{\text{estimated}} = [\boldsymbol{o}_1|\boldsymbol{o}_2|\cdots|\boldsymbol{o}_{N_i}]$$

Consequently, leveraging the loss function defined in **Eq. (3)** enables the rapid, data-free optimization of the multi-wavelength spatially incoherent diffractive processor.

### Model of the digital ANN

The digital decoder ANN employed for image reconstruction is a three-layer fully connected network featuring two hidden nonlinear layers. This network takes wavelength-encoded single-pixel intensity signals as input and produces the reconstructed target image. Specifically, the two hidden layers consist of 128 and 256 neurons, respectively. The mathematical operation for each layer within this reconstruction network is formulated as follows:
$$z_{k+1} = \text{BN}\{\text{ReLU}[\text{FC}(z_k)]\}$$

where $z_k$ and $z_{k+1}$ denote the input and output of the $k$-th layer, respectively. Here, $\text{FC}(\cdot)$ represents a fully connected layer, $\text{ReLU}[\cdot]$ denotes the Rectified Linear Unit activation function, and $\text{BN}\{\cdot\}$ stands for the batch normalization layer.

**Quantitative evaluation metrics**

To quantitatively evaluate the performance of a model, we computed multiple metrics for each linear transformation, including MSE, PSNR, and CosSim. In addition, we assessed the image quality results using MSE, PSNR, and SSIM. For clarity, we also provide brief definitions of these quantitative evaluation metrics:

$$\text{MSE}(A, B) = \text{mean}((A - B)^2) \tag{7}$$

$$\text{PSNR}(A, B) = 10 \cdot \log_{10}\left(\frac{\text{MAX}^2}{\text{MSE}(A, B)}\right) \tag{8}$$

$$\text{CosSim}(A, B) = \frac{A \cdot B}{\|A\| \, \|B\|} \tag{9}$$

$$\text{SSIM}(A, B) = \frac{(2\mu_A \mu_B + C_1)(2\sigma_{AB} + C_2)}{(\mu_A^2 + \mu_B^2 + C_1)(\sigma_A^2 + \sigma_B^2 + C_2)} \tag{10}$$

In these equations, A denotes the target linear transformation or target image ground truth, while B represents the estimated linear transformation or the reconstructed image. In **Eq. (8)**, MAX denotes the maximum pixel value of the image or the linear transformation, set to MAX = 1 in our study. In **Eq. (10)**, $\mu_A$ and $\mu_B$ denote the mean values of A and B, respectively; $\sigma_A$ and $\sigma_B$ are their corresponding standard deviations; and $\sigma_{AB}$ represents the covariance between A and B. The constants $C_1$ and $C_2$ are introduced to ensure numerical stability of the division when the denominator approaches small values.

**Experimental setup**

The experimental configuration is detailed in **Fig. 5b**. We utilized a customized LED array as the incoherent light source, comprising five types of LEDs (Chanzon® 5 mm) with the central wavelengths of 505 nm, 523 nm, 605 nm, 623 nm, and 660 nm. This LED array was modulated by an Arduino UNO R3 microcontroller board to achieve sequential, cyclic illumination across the different wavelengths. The multispectral illumination was collimated and directed onto a DMD (Texas Instruments DLP9500, 10.8 μm pixel pitch), which served as an object display. The DMD displayed 3 × 3 binary images, where each target pixel was represented by a 40 × 40 super-pixel. The target image field was then relayed and demagnified by a factor of 10 to an intermediate image plane using a 4f system composed of lenses L1 (focal length = 500 mm) and L2 (focal length = 50 mm). Subsequently, the optical field propagated through free space for a distance of d1 = 5 mm, was modulated by the diffractive layer, and then propagated another distance of d2 = 12.4 mm before being captured by a CMOS camera (Blackfly BFLY-PGE-12A2M, 1280 × 960 resolution with 3.75 × 3.75 μm pixels). The wavelength-dependent single-pixel intensity signal was obtained by cropping and summing a 7 × 7-pixel central region from the CMOS image.

For the experimental setup, the diffractive layer was quantized to a 4-bit phase depth and featured an array of 90 × 90 elements (totaling 198 × 198 μm in physical dimensions), with each element measuring 2.2 × 2.2 μm. The layer was fabricated using a two-photon polymerization-based 3D printer (Nanoscribe GmbH Quantum X shape). Specifically, using a 63x/1.4 NA objective lens, we achieved a 2.2 μm feature size in a 20-minute printing process. Additionally, a marking box centered around the diffractive layer—as depicted in **Fig. 5f**—was printed to facilitate experimental alignment. Following the initial polymerization, the substrate was transferred to a propylene glycol monomethyl ether acetate (PGMEA) bath for 5 minutes of development, and then immersed in an isopropyl alcohol (IPA) bath for 2 minutes. To prevent structural deformation, a final 1.5-minute rinse was performed using Novec 7100 (methoxy-nonafluorobutane; HFE-7100), an engineered fluid with a surface tension coefficient significantly lower than that of IPA. The random diffuser used in the control experiment was fabricated using the same process described above, except that its thickness was randomly distributed within the range of [0.08, 1] µm.

**Training-related details**

The diffractive optical networks and their corresponding digital decoder ANNs were simulated and trained using the PyTorch framework (v2.3.1). Throughout all model training procedures, we used the Adam optimizer with its default parameters to maintain consistency across models. For the diffractive optical networks, the data-free optimization was conducted over 3000 epochs. The initial learning rate was set to 0.05 and subsequently decayed by a factor of 0.7 at epochs 100, 300, 600, 1000, 1500, and 2100. Conversely, the training of the digital ANN was performed for 100 epochs with an initial learning rate of 0.001, applying a step decay by a factor of 0.7 every 30 epochs. For the diffractive networks, the configuration from the final epoch was utilized. For the digital ANN, however, the best-performing model was selected based on optimal imaging performance evaluated on the validation set. All network training and optimization procedures were accelerated using an NVIDIA GeForce RTX 3090 GPU. In our simulation, the training of a typical digital ANN takes about 12 minutes. A typical data-free optimization of the diffractive processor takes ~15 minutes.

**Supplementary Information includes:**

- **Supplementary Fig. S1**
- **Model of a Multi-wavelength Single-Pixel Diffractive Network**

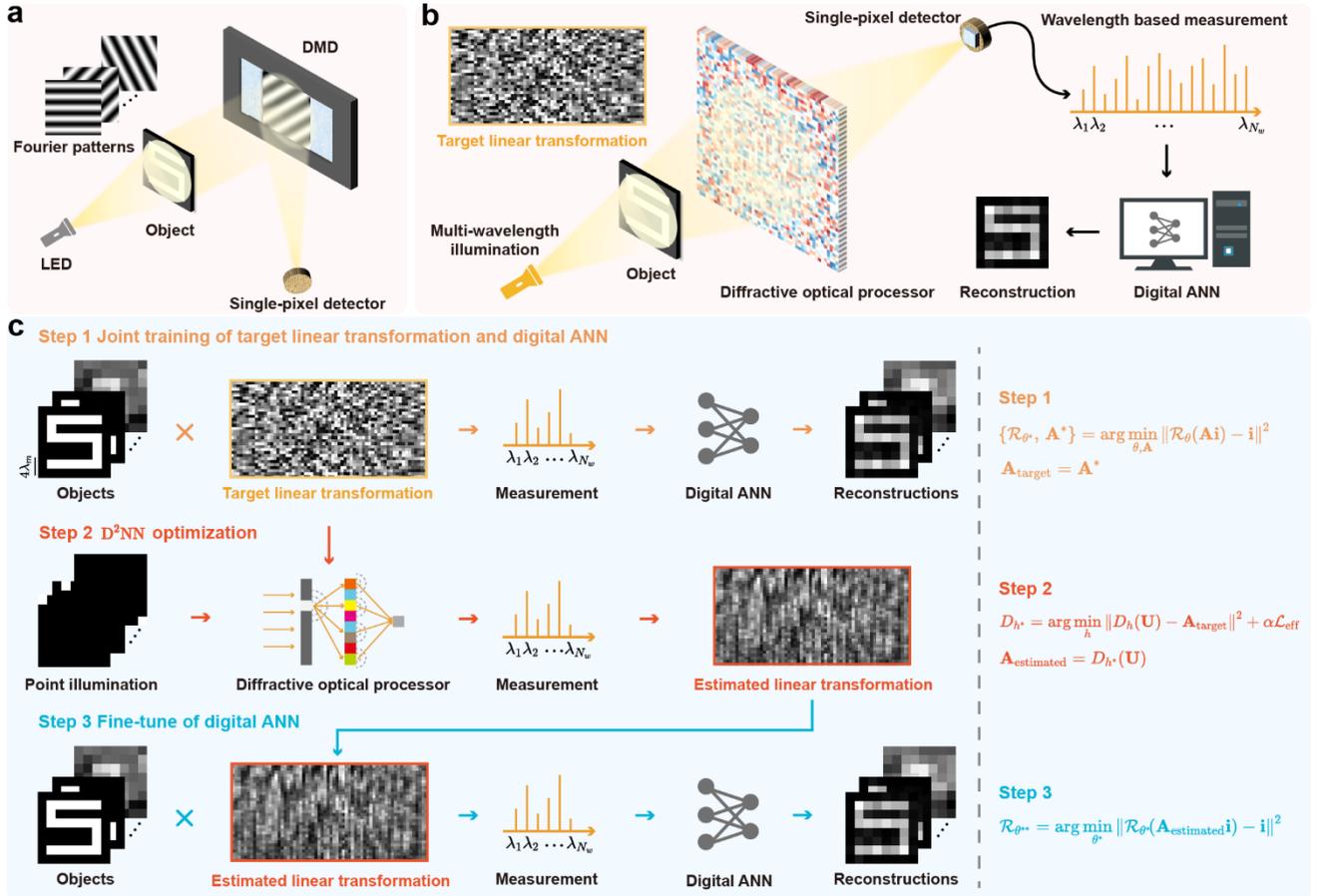

**Fig. 1 Schematics of the wavelength-multiplexed spatially incoherent diffractive optical processor for compressive SPI.** **(a)** Conventional SPI setup utilizing a spatial light modulator, with orthogonal Fourier basis vectors shown as example modulation patterns. DMD, digital micromirror device. **(b)** Working principle of the proposed spatially incoherent diffractive optical processor with multi-wavelength illumination. A target object, e.g., '5' (8×8 pixels), illuminated by a broadband source, is spectrally encoded into $N_w = 32$ wavelength channels by the optimized and fixed incoherent diffractive processor and measured by a single-pixel detector. The measured spectral signal is then decoded by a shallow digital ANN to reconstruct the target image. **(c)** The three-stage training strategy. Step 1: Joint optimization of the target linear transformation ($\mathbf{A}_{\text{target}}$) and the digital ANN. Step 2: Optimization of the diffractive neural network to approximate $\mathbf{A}_{\text{target}}$, producing the optically estimated transformation matrix $\mathbf{A}_{\text{estimated}}$. Step 3: Fine-tuning the digital ANN using the frozen $\mathbf{A}_{\text{estimated}}$ to ensure robust image reconstruction.

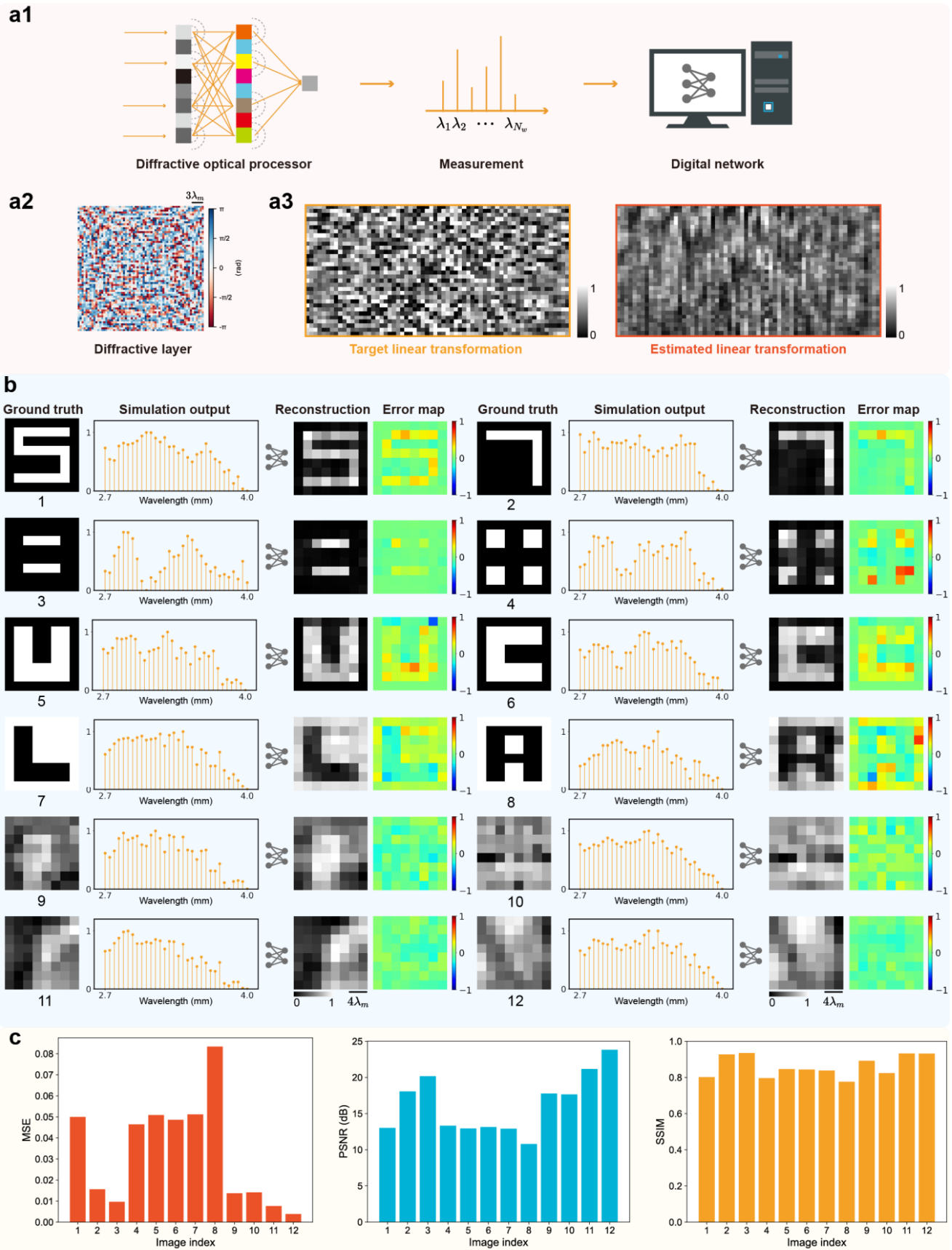

**Fig. 2 Optimized diffractive processor and its imaging performance on test data.** (a1) The inference pipeline: test targets are optically encoded into power spectra by the spatially incoherent diffractive network and digitally reconstructed by a jointly trained shallow ANN. (a2) Phase profile of the optimized diffractive layer. (a3) Comparison between the target linear transformation (derived in Step 1) and the estimated optical transformation (derived in Step 2). Both matrices have dimensions of $32 \times 64$. The quantitative similarity between them is evaluated using: CosSim = 0.93, MSE = 0.04, and PSNR = 13.50 dB. (b) Representative test imaging results for selected samples (never used in the training phase). Diverse image targets encompass digits, resolution patterns, the letters "U," "C," "L," and "A", and other images. Sample indices are denoted beneath the ground-truth images. Across the entire set of 50 test samples, the model achieved average quantitative metrics of MSE = 0.04 ± 0.03, PSNR = 15.82 ± 4.51 dB, and SSIM = 0.83 ± 0.11. (c) Quantitative bar plots of MSE, PSNR, and SSIM for the representative samples shown in (b).

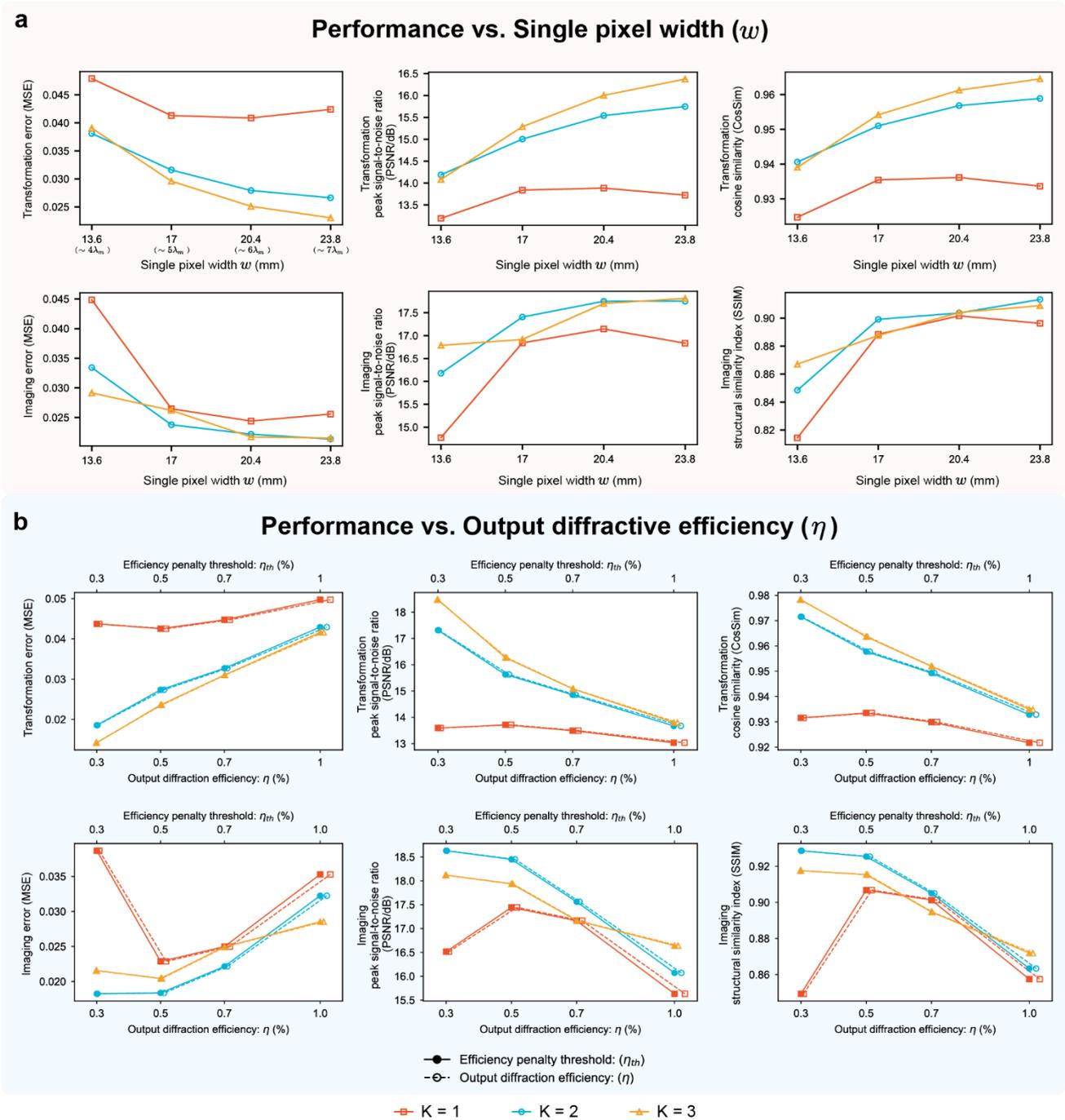

**Fig. 3 Impact of the single-pixel width ($w$) and the output diffraction efficiency ($\eta$) on model performance.** (a) Quantitative metrics for the optical linear transformation accuracy (top row: MSE, PSNR, CosSim) and the final ANN-decoded image quality (bottom row: MSE, PSNR, SSIM) plotted against the single-pixel width ($w$). Markers distinguish the number of diffractive layers: 1 (red squares), 2 (blue circles), and 3 (yellow triangles). (b) A corresponding analysis evaluating the impact of the output diffraction efficiency on both the optical transformation fidelity and the final reconstructed image quality.

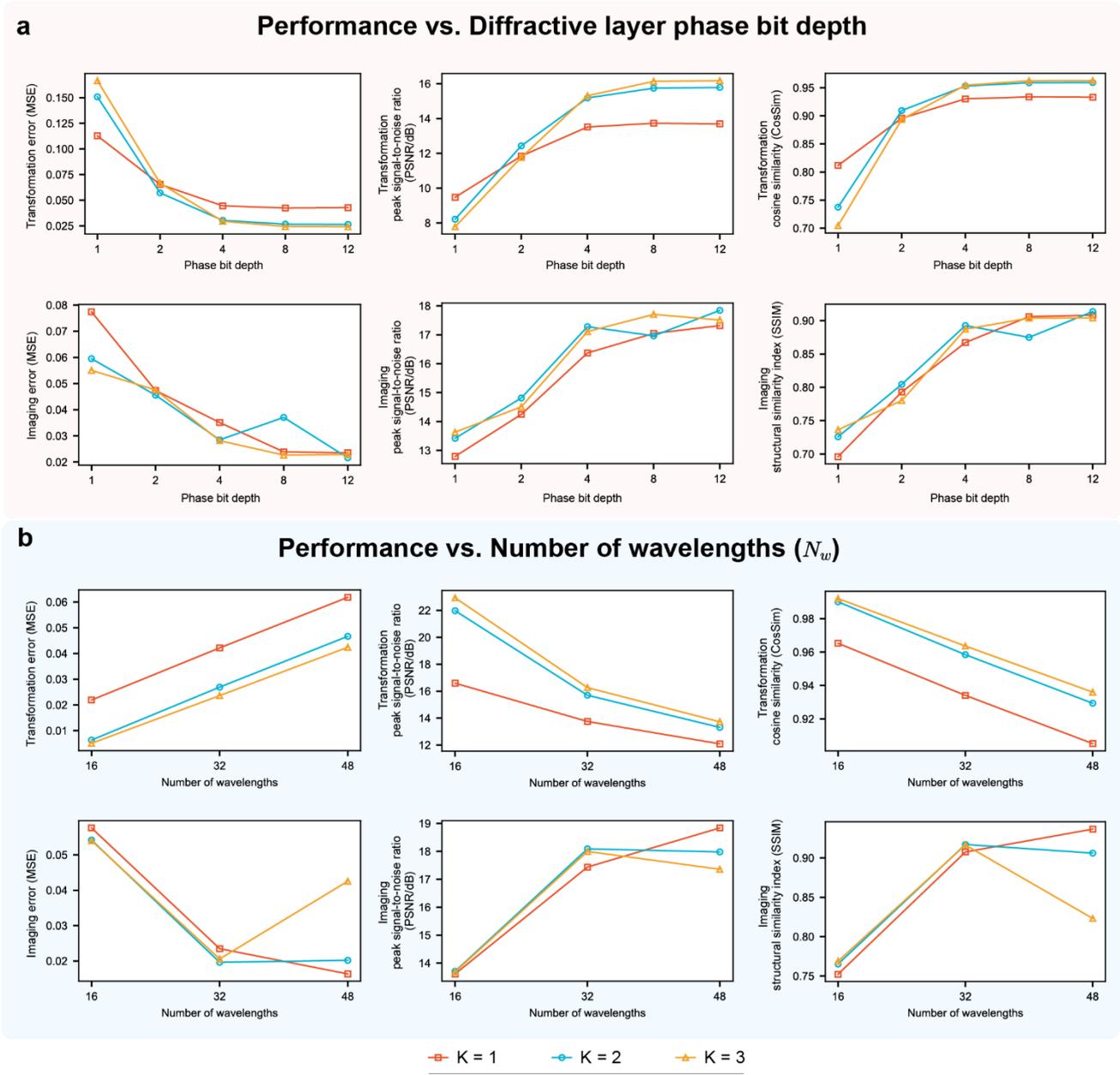

**Fig. 4 Impact of the diffractive layer phase bit depth and the number of multiplexed wavelengths ($N_w$) on model performance.** (a) Quantitative metrics for the optical linear transformation accuracy (top row: MSE, PSNR, CosSim) and the final ANN-decoded image quality (bottom row: MSE, PSNR, SSIM) plotted against the diffractive layer phase bit depth. Markers distinguish the number of diffractive layers: 1 (red squares), 2 (blue circles), and 3 (yellow triangles). (b) A corresponding analysis evaluating the impact of the number of multiplexed illumination wavelengths on both the optical transformation fidelity and the final imaging performance.

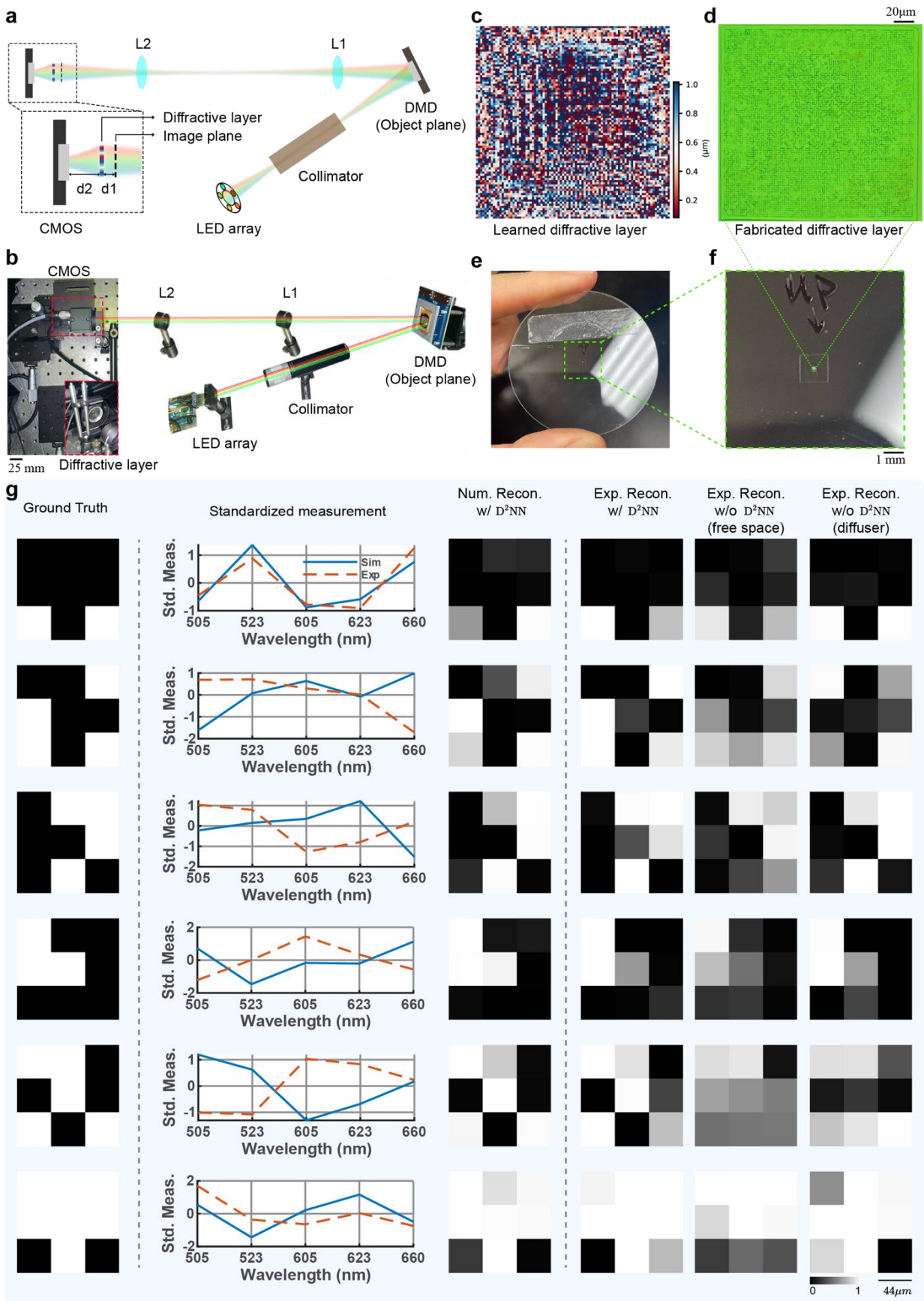

**Fig. 5 Proof-of-concept experiment of a wavelength-multiplexed spatially incoherent diffractive processor.** (a, b) Schematic of the experimental setup. The illumination LED array features central wavelengths of 505, 523, 605, 623, and 660 nm. A DMD (digital micromirror device) is used to display test objects; L1 (500 mm) and L2 (50 mm) form a 4f imaging system. d1 = 5 mm represents the distance from the 4f image plane to the diffractive layer, and d2 = 12.4 mm is the distance from the diffractive layer to the CMOS (complementary metal-oxide-semiconductor) sensor. (c) Optimized thickness distribution of the diffractive layer. (d) Micrograph of the lithographically fabricated diffractive layer. (e, f) The fabricated diffractive layer and its magnified view. (g) Numerical simulation results and experimental imaging results of the diffractive processor for some test images, also compared with the experimental results of two control groups without a D²NN (i.e., free-space propagation and a random diffuser); also see **Table 1** for a quantitative summary of these results.

**Table 1: Average quantitative image quality metrics across different configurations**

| Case | MSE | CosSim | PSNR (dB) | SSIM |
|---|---|---|---|---|
| Num. Recon. w/ D²NNs | **0.0201** | **0.9757** | **17.3962** | **0.9405** |
| Exp. Recon. w/ D²NNs | 0.0258 | 0.9694 | 16.9713 | 0.9247 |
| Exp. Recon. w/o D²NNs (free-space) | 0.1129 | 0.8583 | 10.0047 | 0.6100 |
| Exp. Recon. w/ D²NNs (random diffuser) | 0.0853 | 0.9097 | 13.8005 | 0.7644 |